\documentclass[fleqn,10pt]{wlscirep}
\title{Unveiling Spatial Epidemiology of HIV with Mobile Phone Data}
\usepackage{url}
\usepackage{subcaption}

\author[1,*,+]{Sanja Brdar}
\author[2,+]{Katarina Gavri\'c }
\author[1,3]{Dubravko \'Culibrk}
\author[1]{Vladimir Crnojevi\'c }
\affil[1]{Faculty of Technical Sciences, University of Novi Sad, Novi Sad, 21000, Serbia}
\affil[2]{Institute of Computer Science, University of Heidelberg, Heidelberg, 69117, Germany}
\affil[3]{Department of Information Engineering and Computer Science, University of Trento, Trento, 38122, Italy}

\affil[*]{corresponding author: brdars@uns.ac.rs}

\affil[+]{these authors contributed equally to this work}

\begin{abstract}
An increasing amount of geo-referenced mobile phone data enables the identification of behavioral patterns, habits and movements of people. With this data, we can extract the knowledge potentially useful for many applications including the one tackled in this study - understanding spatial variation of epidemics. We explored the datasets collected by a cell phone service provider and linked them to spatial HIV prevalence rates estimated from publicly available surveys. For that purpose, 224 features were extracted from mobility and connectivity traces and related to the level of HIV epidemic in 50 Ivory Coast departments. By means of regression models, we evaluated predictive ability of extracted features. Several models predicted HIV prevalence that are highly correlated ($>0.7$) with actual values.  Through contribution analysis we identified key elements that impact the rate of infections. Our findings indicate that night connectivity and activity, spatial area covered by users and overall migrations are strongly linked to HIV. By visualizing the communication and mobility flows, we strived to explain the spatial structure of epidemics. We discovered that strong ties and hubs in communication and mobility align with HIV hot spots.
\end{abstract}
\begin{document}

\flushbottom
\maketitle
%
%
\thispagestyle{empty}

\section*{Introduction}

HIV has a devastating social, demographic, and economic effect on Africa \cite{Buve-2002,De-2002}. With a 3.7\% of population infected \cite{website-dhs}, Ivory Coast has the highest prevalence rate in West Africa and a generalized epidemic \cite{Kalipeni-2012}\cite{website-unaids}. This epidemic, where the disease spreads out of the risk groups and affects general population, demands the development of national HIV-prevention plans. Although the prevalence rate appears to have remained relatively stable over the past decade, and is even decreasing, due to prevention of mother-to-child transmission, there is still much work to be done to improve the health system to enable a more effective response to HIV. Deeper understanding of the epidemics can help find ways to suppress HIV further and modern technologies that deal with human mobility phenomena may help respond to that challenge. 

Mobile phone communication engendered the era of big data by creating huge amounts of call detail records (CDRs). Cell phone service providers collect these records whenever a phone is used to send a text message or make a calls. These records contain the time of the action, identifiers (IDs) of sender, receiver and the cell towers used to communicate. In this way, mobile phones provide approximate spatio-temporal localization of users and create an immense resource for the analysis of human mobility and behavioral patterns\cite{Becker-2013, Candia-2008, Wesolowski-2010}. In a burst of new applications built on mobile phone data\cite{Blondel-2015}, we emphasize those of great practical importance such as urban planning\cite{Becker-2011}, disaster management\cite{Bengtsson-2011}, transportation mode inference\cite{Wang-2010}, traffic engineering\cite{Caceres-2012}, deriving poverty indicators\cite{Smith-2014} and crime prediction\cite{Bogomolov-2014}. 

Currently, there is a growing interest in the mining of mobile phone data for epidemiological purposes\cite{Lima-2013,Tizzoni-2014}. Mining can advance research in epidemiology by shedding light on relationships between disease distribution, spread and incidence on one side, and migrations, everyday movements and connectivity of people on the other side. Up to now, only a few studies have used mobile phone data to quantify those relationships based on real disease distribution data. Wesolowski and co-workers explored the impact of the human mobility to the spread of malaria\cite{Wesolowski-2012}. They analyzed CDR data collected by a mobile phone service provider in Kenya over the period of one year and discovered how human mobility patterns contribute to the spread of the disease beyond what could be possible if it was transferred only by insects. Another study carried by Martinez et al.\cite{Frias-2012} investigated the effect of government alerts during H1N1 flu outbreak in Mexico on the diameter of the mobility of individuals. Bengtsson and co-workers \cite{Bengtsson-2011} estimated population movements from a cholera outbreak area and suggested the use of information obtained for disease surveillance and resolving priority in relief assistance. Those pioneering works usher in the emerging field of digital epidemiology\cite{Salathe-2012}. 

To the best of our knowledge, the study we describe here is the first attempt to use mobile phone data to explore the complex structure of HIV epidemics. Significant scientific effort is aimed at identifying the driving factors of HIV spread. Most frequently mentioned are poverty, social instability and violence, high mobility, rapid urbanization and modernization. The differences among these factors could help explain the spatial disparity observed in prevalence rates. Messina et al. examined geographic patterns of HIV prevalence in Democratic Republic of Congo \cite{Messina-2010}. They showed that spatial factors: the prevalence level in the 25 km range and the distance to the urban areas are strongly connected to the risk of HIV infection. The impact of migration on the spread of HIV in South Africa has been studied in \cite{Coffee-2007}, where authors developed a mathematical model to compare the effects of migration and associated risk behavior. In the early stage of epidemics, migration impacts the HIV progression by linking geographical areas of low and high risk. In the later stage, the impact is mainly through the increase the high-risk sexual behavior. However, the migration in the study was quantified through surveys,  in which the participants were questioned about movement history, and the study included only two migration destinations, limiting both the extent of the study and the quality of data that was used.

Nowadays, when overwhelming amounts of mobile phone data exists, providing us with insight into the movements and activity of millions of people over large areas, we can try to utilize it for new studies of the epidemiology of HIV. In the study described here, we conducted a comprehensive analysis of two data sets offered within the Data for Development (D4D) Challenge \cite{Blondel-2012}. Our research was guided by the following hypothesis: the risk of HIV infection is associated with spatial and behavioral factors that can be detected from the collection of data available. We were particularly interested in tracking population movements and inferring the strength of communication between departments of Ivory Coast with different prevalence rates.

\section*{Results}

\subsection*{Spatial distribution of HIV}
\begin{figure}[h!]
\centering
\includegraphics[width=0.92\linewidth]{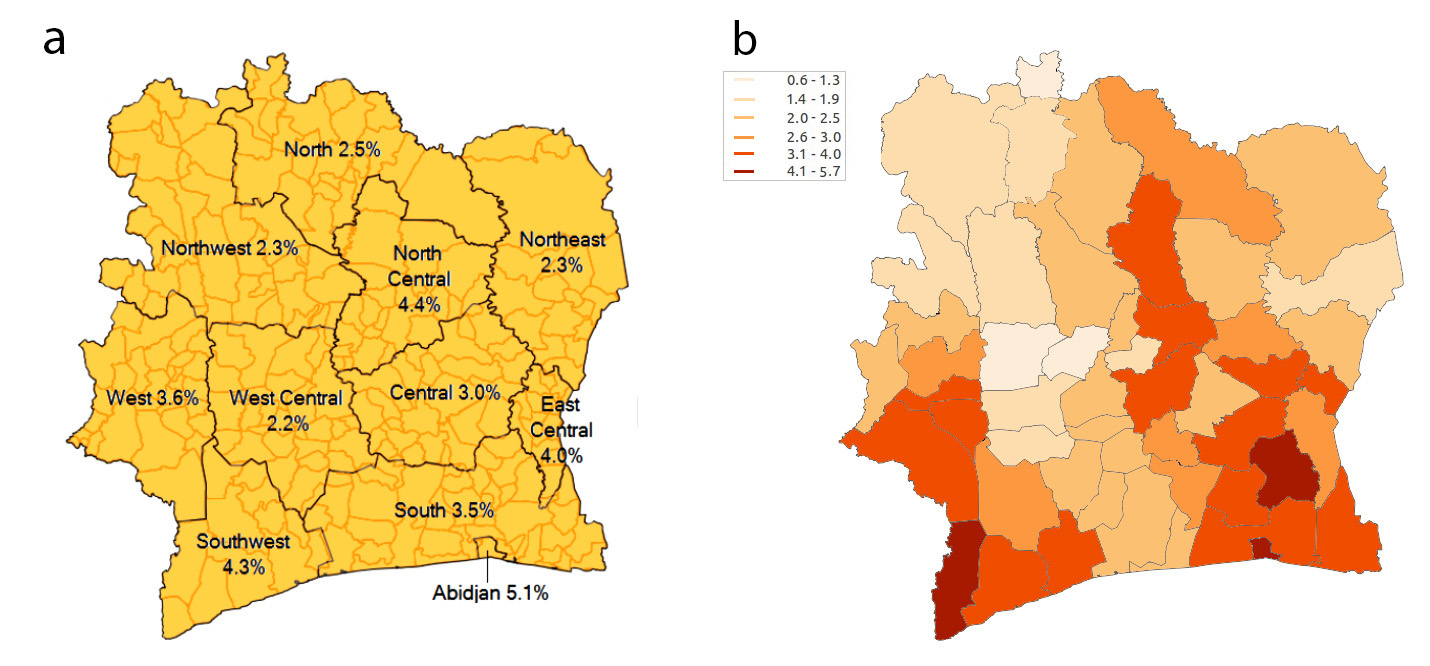}
\caption{(a) HIV prevalence rate by administrative regions - DHS data (b) HIV prevalence rate by departments for 15–49 year-olds population; estimated values range between 0.6 and 5.7\% .}
\label{fig:hiv-distribution}
\end{figure}

To determine the health status of a population, Demographic and Health Surveys (DHS) periodically organizes surveys to gather relevant data, focusing on specific countries. In our study we used the DHS data collected in Ivory Coast during their 2012 campaign~\cite{website-dhs}. Based on the measurement, DHS provides estimates of HIV prevalence at sub-national level, but with a low spatial resolution, determined by 10 administrative regions (Fig. \ref{fig:hiv-distribution} (a)). Estimates of the HIV prevalence range from 2.2 to 5.1\% and reveal the spatial variability of the distribution of HIV-infected across country. Due to initiatives to examine further the spatial heterogeneity of HIV\cite{website-UNAIDS-reference-group}, new methods emerged, aiming to provide HIV estimates at a finer resolution. An approach that employs kernel estimation based on spatial DHS measurements, with an additional adjustment to UNAIDS data, made estimates for 50 departments of Ivory Coast available (see Methods). After redistributing disease frequencies across 50 departments, the HIV prevalence map (Fig. \ref{fig:hiv-distribution} (b)) shows higher spatial variability (from 0.6 to 5.7\%) in the disease distribution. We can notice the hot spots of epidemics – departments severely hit by HIV. The map also enables us to explore links between the connectivity and mobility patterns derived from D4D data and HIV prevalence with increased spatial resolution. Although the quality of HIV estimates (imposed by DHS measurement sampling) at department level varies from good and moderate to uncertain, the data has the highest spatial resolution currently available for studying the HIV epidemic in Ivory Coast. 

\subsection*{Communication and mobility patterns}
Social interactions and mobility mediate the spread of infectious diseases\cite{Read-2008, Tizzoni-2014, Belik-2011}. When examined in a spatio-temporal context, they can uncover how a disease propagates and finally explain the variability in the prevalence distribution. To understand spatial epidemiology of HIV in Ivory Coast better, we analyzed the collective communication and mobility connections at the level of departments. We estimated pairwise connections among sub-prefectures by measuring communication and mobility flows. To accomplish that, we explored the "antenna-to-antenna data" (SET1) and the "long term individual trajectories" (SET3) D4D datasets~\cite{Blondel-2012}.

SET1 provided us with insight into the communication flow between each pair of antennas on an hourly basis. The strength of the communication flow is expressed through the number of calls. We assigned each antenna to a corresponding department and then aggregated the number of calls at the department level during a 5-month observation period.  SET3 shed light on the mobility of people, providing the geographic location of users while using their phone to make calls or send messages. Since records in SET3 contain the user ID, location at the sub-prefecture resolution and time stamps indicating when the phone was used, we were able to use them to estimate the location of the user's home. Based on the the most frequent location, we assigned each user to his/her home department. Then we counted the user's movements from home to other locations over the entire 5-month observation period and aggregated users' movements at the department level.

In the pairwise communication and mobility matrices, obtained in this way, we identified \textit{strong ties} for each department, which represent links to other departments with the connection strength higher than the average (see Methods). Before searching for the strong ties, we normalized the matrices by the corresponding population sizes. SET1 encompasses 5 million of users. We distributed them into departments, using population frequencies provided by Afripop data\cite{Linard-2012}, and used the per-department populations obtained to normalize the communication flows.  To normalize the migration flows, we used estimates based on the derived home locations of the users to calculate the required population size per department. Each communication or mobility flow was normalized by the corresponding population size of originating department. The overall flow between two departments was then quantified as sum of normalized flows in both directions. This enabled us to eliminate the bias caused by the different population sizes when identifying the strong links. 

\begin{figure}[h!]
	\centering
	\includegraphics[width=0.95\linewidth]{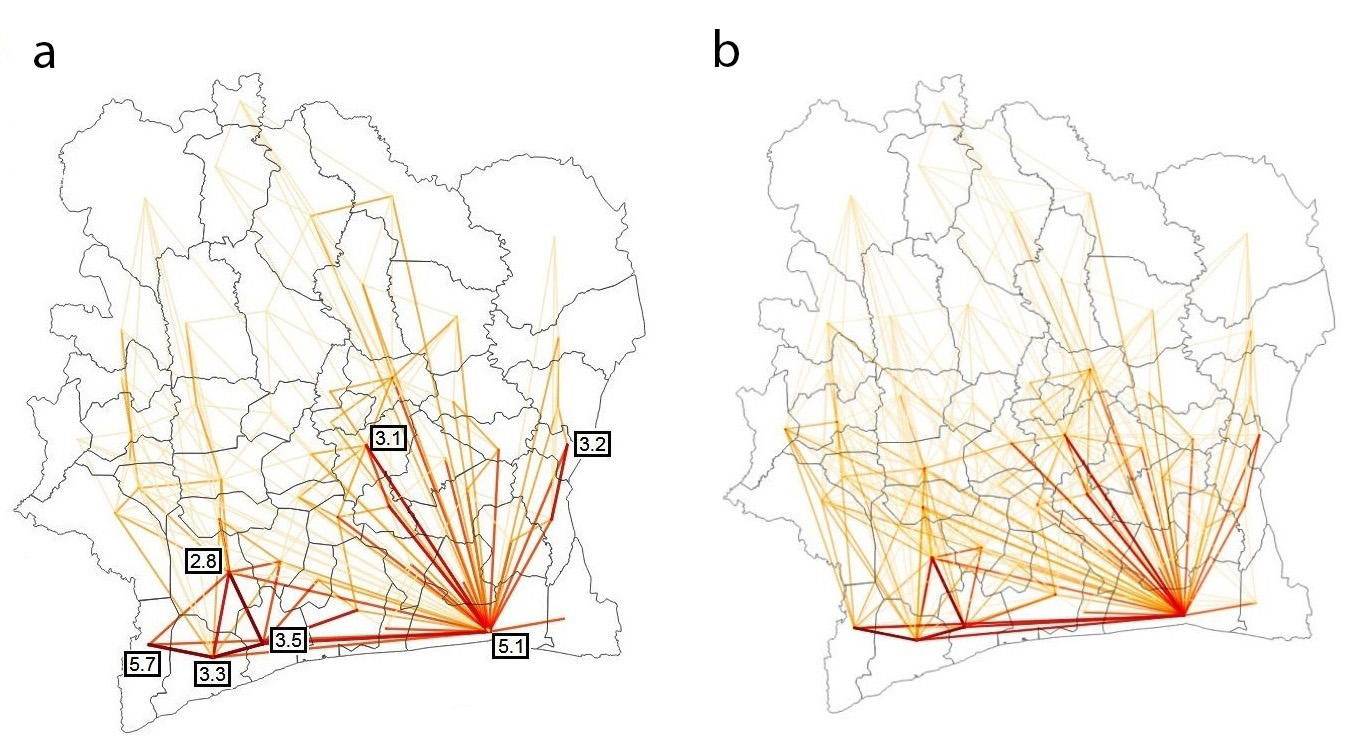}
	\caption{Strong connectivity ties for (a) overall communication (b) night communication. The hubs are labeled with the corresponding HIV prevalence rate shown in Fig. \ref{fig:hiv-distribution} (b). Link thickness and color, ranging from yellow to red, are proportional to the strength of communication flow.}
	\label{fig:sties-connectivity}
\end{figure}

The strong ties discovered in communication flows are shown in Fig. \ref{fig:sties-connectivity} (a). This visualization emphasizes the strongest links further and communication hubs emerge. Remarkably, the hubs correspond to HIV hot spots and we can also notice that larger hubs have higher prevalence rates. Additionally, we visualized the night communication, constrained to the time interval between 1AM and 5AM, and obtained a similar structure of the connectivity graph - Fig. \ref{fig:sties-connectivity} (b). 

\noindent The links correspond to relative rather than absolute flow, which we obtained by dividing the flow with the maximum value of flow in the set of strong ties. In both graphs we can notice how departments in the north part of the county have weaker links and this may explain why they have smaller HIV prevalence.

\begin{figure*}[h!]
\centering
\includegraphics[width=0.95\linewidth]{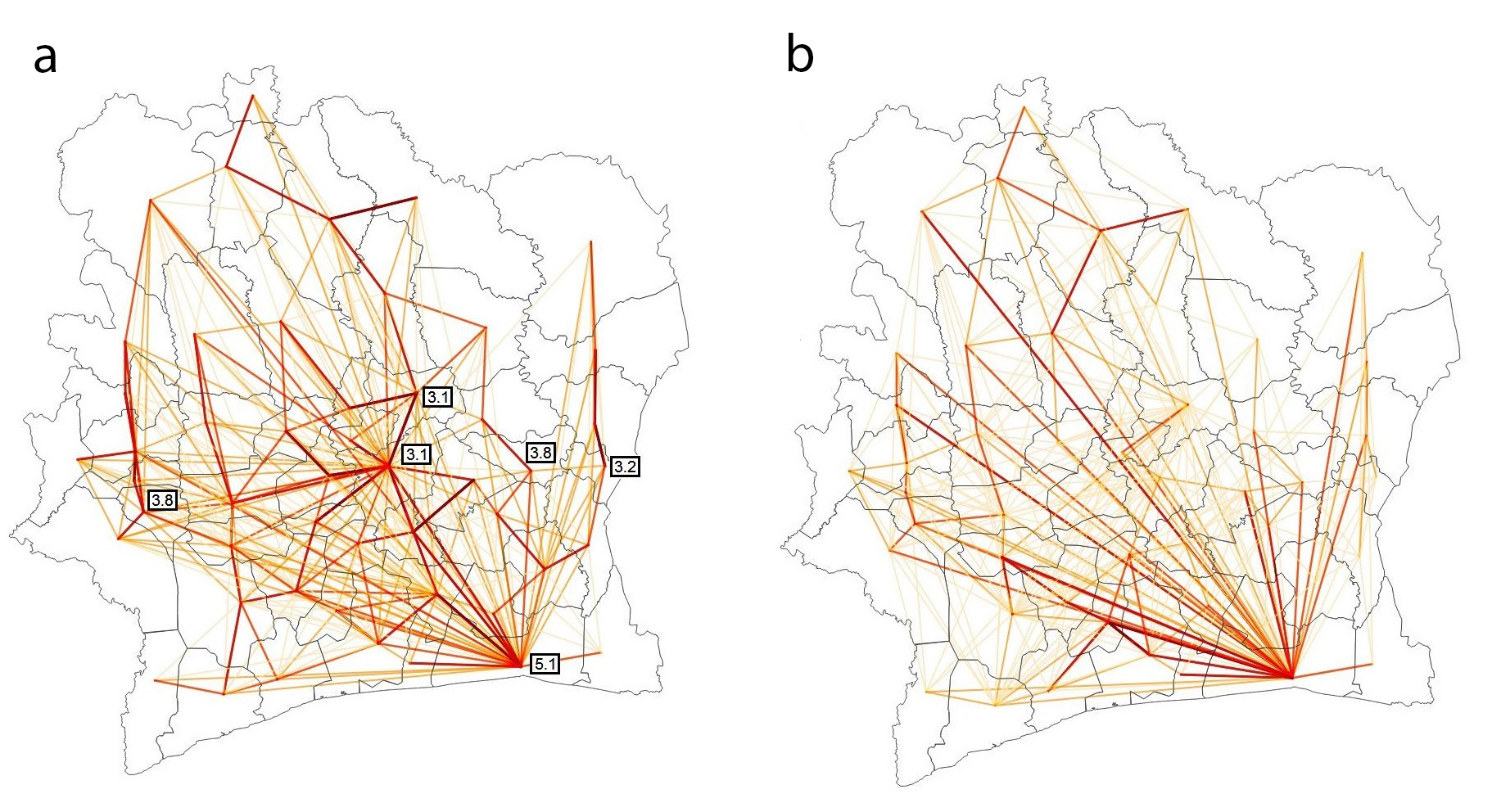}
\caption{Strong mobility ties discovered through summarizing (a) all mobilities (b) mobilities with 3 days or longer spent at the destination. The hubs are labeled with the corresponding HIV prevalence rates shown in Fig. \ref{fig:hiv-distribution} (b). The link thickness and color, ranging from yellow to red, are proportional to the strength of mobility flow.}
\label{fig:mobility}
\end{figure*}

The strong ties discovered in mobility flows (Fig. \ref{fig:mobility} (a)) have an obvious localized character. They connect the departments that are geographically close, but, on a global scale, we can also observe strong migratory pathways. One connects the two largest hubs - the largest city Abidjan (5.1\% prevalence rate) and the capital city Yamoussoukro (3.1\% prevalence rate). From the center of country we can notice strong pathways to the region in the West (3.6\% prevalence rate, Fig. \ref{fig:hiv-distribution} (a)) and the North-central region (4.0\% prevalence rate, Fig. \ref{fig:hiv-distribution} (a)). The East-central region, with a prevalence rate of 4.0\% is strongly connected to Abidjan. The map of the mobility flows revealed the pathways that connect regions with higher prevalence. 

In addition to the observed general mobility of users, we explored the long-term mobility. We measured how long users stay at their destinations and in our migration analysis considered only those stays in which the users stayed longer than 3 days. The strong ties discovered in long-term mobility flows are shown in Fig. \ref{fig:mobility} (b). The connectivity graph obtained, reveals how long-term migrations link departments further away. Interestingly, Abidjan emerged as the most prominent hub for those migrations. In this light, we can denote this city, with the largest prevalence rate and high connectivity, as a driver of epidemic in the Ivory Coast. As such, Abidjan needs careful monitoring of mobility flows, especially the high-risk longer-term mobilities, in order to prioritize interventions and control the further spread of HIV.

\subsection*{Extracted features}
For each department of Ivory Coast, numerous features were extracted during the course of the study presented, with the goal to quantify behavioral and mobility patterns potentially relevant to the measured HIV prevalence rate. Overall, we extracted 224 different features and grouped them into 4 categories: connectivity, spatial, migration and activity (phone use).

The connectivity features were obtained from the SET1 data. The communication flow is expressed through the number of calls and their duration in SET1. Using the information of the originating and terminating antenna, for each department, we aggregated its inner, originating, terminating and overall communication. The overall communication was further separated based on the type of day and time of day constraints. We considered two types of days: weekdays and weekends, and used 1-hour time slots (00-01h, 01-02h, ... , 23-24h) and 8-hour time slots (00-08h, 08-16h, 16-24h) to express the time within a day.  For each of these discrete intervals, the features related to the number of calls represent the cumulative sum over the whole five-month observation period. Once extracted they were normalized by the corresponding department population size, estimated based on Afripop data\cite{Linard-2012} and rescaled to fit the 5 million of users monitored in our data set. Features related to the duration of calls represent average values. Overall, 120 connectivity features related to different time slots and type of days were extracted; half to describe the number of calls and half to describe the average duration of calls.

Spatial, migration and activity features were derived from SET3 data. To craft spatial features we explored positions and the distribution of locations visited by users. We measured the radius of gyration, area and the perimeter of convex hull of users' movements, as well as the  diameter of their range~\cite{Gonzalez-2008,Williams-2014,Csaji-2013}. The features were derived both for all locations visited by a user, as well as specific subsets of locations: visited at night, on weekdays, weekends, weekday and weekend nights. In addition, we calculated the total distance traveled by each user. In total, 25 spatial features were created, representing 95 percentile values across users matched to departments based on their home location. Interestingly, we first considered averaged instead of 95 percentile values for users in corresponding departments, but for predictive models  better results are achieved when spatial features capture only the top five percent of users; i.e. the patterns of users that cover larger regions through their mobility have higher predictive power on the prevalence of HIV. 

To extract migration features we tracked the changes in locations. Every time a user changed department, we added a single migration link from his home to the observed department. We summarized all movements into a pairwise migration matrix by iterating this procedure for all users. Beside quantifying all movements, we also identified those where users were away from home for more than defined number of days (1, 2, ..., 10) to explore longer-term migrations. The features were divided further according to the direction of the mobility into ''in'' or ''out'' migration, bringing their total number to 22. 

The activity features were extracted similarly to the connectivity features. However, in SET3, we cannot distinguish the direction of communication (in or out), nor do we have the duration of communication. Therefore, we refer to those features simply as activity since they can count only when and where users were active. As with the connectivity features we considered two types of days: weekdays and weekends. The time of day was again considered in 1-hour time slots, 8-hour time slots and whole days. The total number of activity features used was 57.

All the features capture the cumulative effect of human connectivity or mobility observed over a five-month period. We focused on this long-term perspective in our feature extraction, in order to understand the spatial distribution of HIV prevalence better.

\subsection*{Predictive models}
HIV prevalence rates across the departments of Ivory Coast range from 0.6 to 5.7\%. Each of the 50 departments was represented with a vector of extracted feature values and the corresponding prevalence rate. In this feature space, we built regression models and evaluated their performance when predicting a department's prevalence rate. All features were normalized by dividing each feature with its mean value across the whole data set, before regression was attempted. 

Experiments were conducted using two different regression methods: Ridge\cite{El-2011} and Support vector regression (SVR)~\cite{Gunn-1998}. The regression models were initially built using the four different groups of features separately. In order to select smaller subsets of most relevant features, both regression methods were subjected to recursive feature elimination RFE\cite{Guyon-2002} method. In the final stage, we considered an ensemble approach -- stacked regression~\cite{Breiman-1996} -- through which we fused 4 heterogeneous feature sets, building a single integrated prediction model. 

The prediction of disease levels needs careful evaluation~\cite{Bodnar-2013} in order to avoid situations in which models built on randomly generated data work comparatively well to those created on possibly meaningful data. Therefore, to estimate the predictive capacity of a model, we measured the prediction errors and correlations between the predicted and actual values for the models built on real data and the same models created based on random data sets, obtained by randomly permuting values for each feature.  

Experiments were divided into two parts: the first stage focused on the 15 departments with good and moderate estimates of HIV prevalence, while in the second we used data for all 50 departments. In Tables \ref{tab:predictions-evaluation1} and \ref{tab:predictions-evaluation2}, we report the correlation coefficients ($\rho$) and relative root mean square errors ($RRMSE$) produced by the  models during leave-one-out (LOO) cross-validation, for two experimental setups (15 and 50 departments). 

LOO evaluation enabled us to select the best model among those we built. On the subsample of 15 departments, the models built with SVR, with RFE, perform best. SVR models surpassed Ridge, and reducing the size of the feature set with RFE improved performance of both, but the SVR method benefited more from the RFE procedure than Ridge. The highest correlation coefficient (0.753) between the predicted and actual values is achieved with the SVR on a reduced set of 6 most relevant spatial features. The lowest error of 0.287 is reached by combining regressors learned on different sets of features. Through the linear combination of the four models, the ensemble approach predicts HIV prevalence values that are well correlated with actual ($\rho = 0.710$). All models built on the real features outperformed their random counterparts.

\begin{table*}[h!]
	\caption{Evaluation of predictive models on good and moderate HIV estimates - Correlation coefficient (Relative Root Mean Square Error): $\rho~ (RRMSE)$}
	\centering 
	\begin{tabular}{r|cccc}
		\hline
		& \multicolumn{4}{c}{Predictive models} \\
		Features & Ridge  & Ridge+RFE & SVR & SVR+RFE \\
		\hline 
		Connectivity features (SET1) & 0.624 (0.331) & 0.626 (0.331)  & 0.661 (0.306) & \bf{0.669 (0.301)}\\			
		Spatial features (SET3) & 0.639 (0.434)   & 0.703 (0.376) & 0.544 (0.351) & \bf{0.753 (0.294)}\\		
		Migration features (SET3) &	0.585 (0.369)  & 0.585 (0.369) & 0.678 (0.307) & \bf{0.691 (0.288)} \\ 			
		Activity features (SET3) &	0.618 (0.339)  &	0.645 (0.325)	& 0.633 (0.316) & \bf{0.664 (0.302)}\\
		Ensemble & 0.610 (0.327)  &	0.601(0.327) &	0.659 (0.305) & \bf{0.710 (0.287)}  \\			
		Best Random &	-0.231(0.511)  & -0.066 (0.480) & -0.065 (0.479) & 0.070 (0.441)\\	
		\hline
	\end{tabular}
	\label{tab:predictions-evaluation1}
\end{table*}

The second part of the experiments evaluated the proposed methods and extracted features on the full set of 50 departments, including those with uncertain estimates on HIV. Table \ref{tab:predictions-evaluation2} reports the obtained results. As expected the performance declined. Predictions are moderately correlated with actual values. The best result $\rho = 0.627$, $RRMSE$ = 0.509 is achieved with the SVR model on a reduced subset of activity features. Ensemble approach that combines four SVR+RFE models results in $\rho = 0.518$ and $RRMSE = 0.514$. Still, the models created on randomly permuted features predict HIV with higher errors and without correlation with actual values and, thus, underperform those built on real features.

\begin{table*}[h!]
	\caption{Evaluation of predictive models on all HIV estimates - Correlation coefficient (Relative Root Mean Square Error): $\rho~ (RRMSE)$}
	\centering 
	\begin{tabular}{r|cccc}
		\hline
		& \multicolumn{4}{c}{Predictive models} \\
		Features & Ridge  & Ridge+RFE & SVR & SVR+RFE \\
		\hline 
		Connectivity features (SET1) & 0.467 (0.556) & 0.481 (0.546)  & 0.501 (0.516) & \bf{0.508 (0.514))}\\			
		Spatial features (SET3) & 0.363 (0.540) & \bf{0.431 (0.523)} & 0.310 (0.552)  & 0.336 (0.545)\\		
		Migration features (SET3) &	0.269 (0.630)  & 0.315 (0.613)  & 0.291 (0.637) & \bf{0.375 (0.599)}\\ 			
		Activity features (SET3) & 0.511 (0.542)  & 0.542 (0.535) & 0.522 (0.537) & \bf{0.627 (0.509)} \\
		Ensemble & 0.500 (0.527) & {\bf{0.543}} (0.519) & 0.535 (0.515) & 0.518 \bf{(0.514)}  \\			
		Best Random & 0.020 (0.760)	& 0.202 (0.657) & 0.139 (0.630) & 0.038 (0.607) \\	
		\hline
	\end{tabular}
	\label{tab:predictions-evaluation2}
\end{table*}

\subsection*{Feature contribution}

Once a regression model is built, we can use it to estimate the risk of disease in defined spatial units. Furthermore, we can examine what the model learned from the data. Model explanation techniques \cite{Strumbelj-2011, Strumbelj-2014} can unveil black-box predictive models by estimating contributions of each feature over the whole range of its input values. For example, we can examine how changes in an activity feature affect the value of the HIV prevalence rate, obtained by the model built. The outcome is a plot of the contribution as a function of feature values. This model-explanation procedure provides us with the opportunity to identify specific features that impact prevalence rate most of all and to quantify their contribution. The features identified in this manner can later be continuously measured and leveraged for the monitoring of changes in the HIV prevalence rate and to create early warning signs for possible increase of the infected population. 

\begin{figure}[h!]
	\centering
	\includegraphics[width=\linewidth]{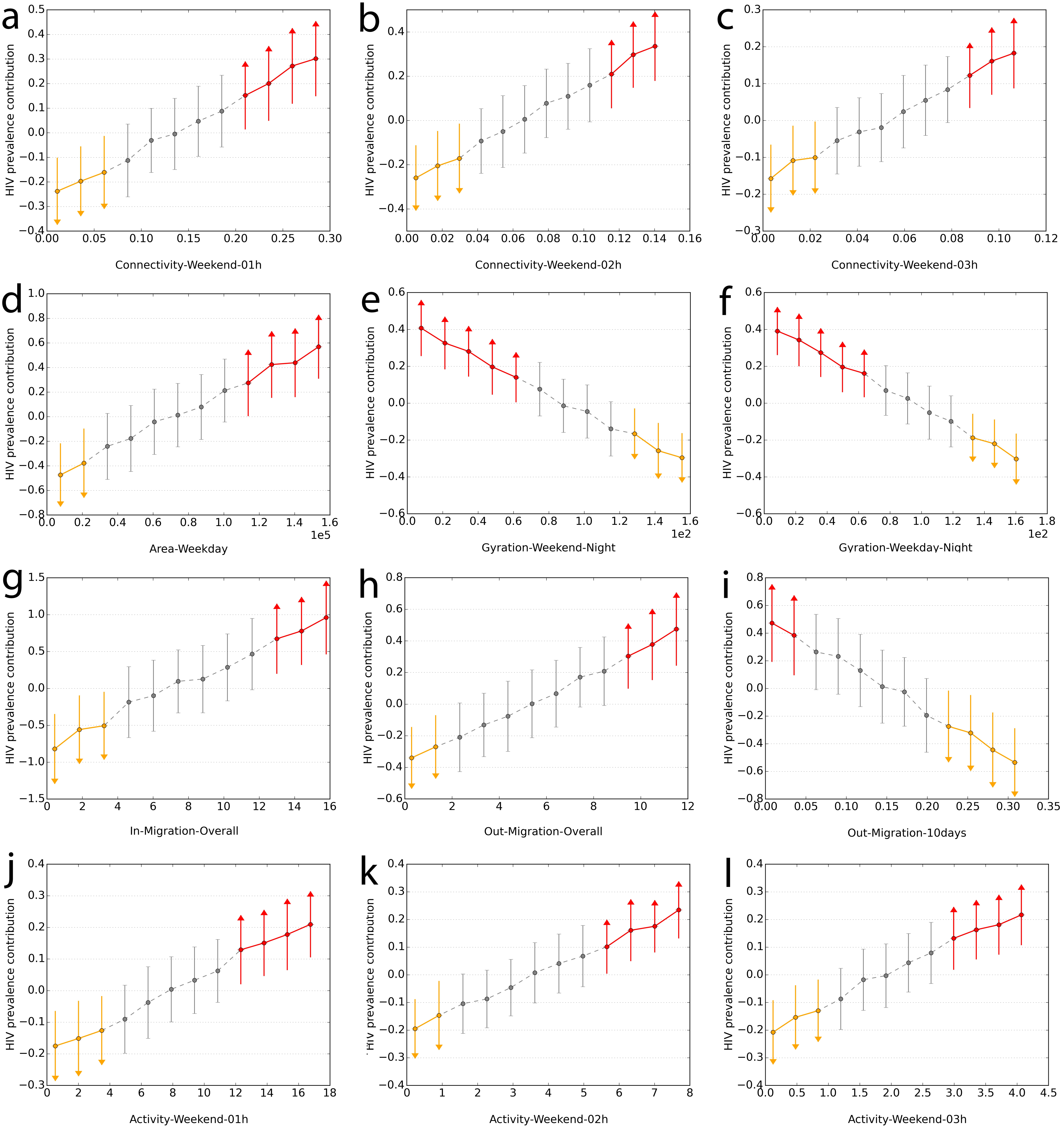}
	\caption{Feature contribution graphs for 12 features; top 3 features for 4 types of features. Points correspond to the mean contribution and error bars correspond to standard deviation. Red color indicates strong association to higher HIV prevalence, and orange to lower HIV prevalence.}
	\label{fig:contribution}
\end{figure}

To conduct the feature contribution analysis, we used the best model (SVR+RFE) built for each set of features, since the ensemble method is just an additive combination of models built on different sets. In the analysis we used models built on a subsample of departments (15 with good or moderate HIV estimation) and focused on the top 3 features, selected by running the RFE procedure until only 3 features remain. The remaining features have highest impact on HIV prevalence prediction. For the selected features ($f_{t,i}$, where $t$ denotes set of features and $i$ is index of feature in that set) we conducted contribution analysis. We calculated the contribution for each feature over the full range from its minimal to maximal value in $m$ equally distributed points. The contribution analysis included the randomization process to create two instances as inputs to regression model. The first instance is a vector where each feature value is sampled at random from the data set $t$. The second instance differs in $i^{th}$ feature which is not random but takes a particular value from set of previously defined $m$ values that are currently under contribution analysis. The contribution of the feature is the difference between the outputs of the regression model produced using the first and the second instance as input. Due to the randomization process this procedure is repeated for a defined number of iterations. By averaging the results from all iterations, we obtained the final value for contribution. In addition to this value, we also report the standard deviation of the values obtained in each iteration, which provides information on the contribution stability and quantify complex interactions among features. We created plots (Fig. \ref{fig:contribution}) for 12 features - top 3 for each of four data sets, sampled in $m=12$ points with contributions calculated through 100 iterations. In addition, the 12 graphs that correspond to features ranked from $4^{th}$ to $6^{th}$ place for each data set are provided in the supplement - Fig. S2. All graphs contain points of the mean contribution and error bars in the length of standard deviation. Red color indicates points with feature values that are associated with increased HIV prevalence, and orange color indicates feature values that are associated with decreased HIV prevalence. The gray part of graph denotes the range where the standard distribution crosses zero, meaning that contribution is neither strongly positive nor negative.

Contributions of the three connectivity features are presented in Fig. \ref{fig:contribution} (a), (b) and (c). Top three features represent the communication flow expressed as the number of calls per resident of a department during the days of weekend in the time slots 01-02 AM, 02-03 AM and 03-04 AM, over a 5-month period. We can notice that the top connection features are related to weekend night-time communication and all have a positive slope. A similar graph (Fig. S2) is obtained for the $5^{th}$ ranked feature related to weekday 03-04 AM communication. According to the model, the departments with higher night-time communication have a higher prevalence rate. In further analysis of the contribution plot shown in Fig. \ref{fig:contribution} (a), values higher than 0.2 can be seen as indicators of behavior increasing the risk of infection and thus critical for HIV. For example, for the department where this feature has the maximum value, the expectation of HIV prevalence is by 0.3  $\pm$ 0.15 higher than average. The plots for features ranked at $4^{th}$ and $6^{th}$ place (Fig. S2), refer to average call duration during the hours of early morning (06-07 AM) and contribute to HIV prevalence in a different manner. Those graphs have a negative slope, indicating that, for departments were people have longer talks early in the morning, we can expect lower HIV prevalence. We can observe this as a social signature \cite{Saramaki-2014} and may hypothesize that longer talks early in the morning could be an indicator of emotionally close relationships and lower-risk behavior. 
 
In  the contribution analysis of spatial features, area and gyration stand out as features with higher impact. Area is measured over weekdays and gyration over weekday and weekend nights. The model suggests that departments were people, through their overall movements, tend to cover a larger area,  have a higher HIV prevalence rate (Fig. \ref{fig:contribution} (d)). This is also confirmed by the $4^{th}$ ranked feature, which measures the area covered over weekends (Fig. S2 ). On the contrary, gyration, a measure of standard deviation from the mean location, negatively impacts HIV (Fig. \ref{fig:contribution} (e),(f) and also Fig. S2). But it is no surprise that small gyration indicates higher HIV, since it has already been shown in other studies that there is a higher expectation of shorter movements in the denser urban areas\cite{Noulas-2012}, and those urban areas are usually more affected by HIV. Interestingly, when the area covered is tracked only during the hours of the night, the contribution graph has a negative slope as it does in the case of gyration (see graph for $5^{th}$ ranked feature - area covered during weekday nights, Fig. S2).

The contributions of overall in and out migration features are shown in Fig. \ref{fig:contribution} (g), (h). Both plots indicate that larger migration flows are associated to higher HIV prevalence. We can notice the strong impact of incoming  migrations: for the department where this feature has the maximum value, the expectation of the HIV prevalence is by 1.0  $\pm$ 0.5 higher than the average. Among the top three features is the one that quantifies the number of out migrations per resident of a department, with the time clause of staying for more than 10 days. Its contribution plot, presented in Fig. \ref{fig:contribution} (i), shows negative impact. 
The plots for features ranked between $4^{th}$ and $6^{th}$ place (Fig. S2) further show that out migrations, with stays longer than one day have a positive slope, and those with stays longer than 5 or 9 days exhibit a negative slope. The contribution analysis of the migration features uncovers an interesting phenomenon - the overall amount of migrations is linked to higher HIV prevalence, and this positive slope remains true for migrations up to a few days, but beyond that, the slope becomes negative. The slope changes once the thresholds of 4 days for out migrations and 3 days for in migrations are reached. Thus, the model built suggests that the risk comes from shorter stays at host departments and higher dynamics in migrations, while the longer stays are associated with lower HIV prevalence.

The contribution of the activity features, expressed through the number of calls and SMSs per residents of a department, are shown in Fig. \ref{fig:contribution} (j), (k), (l). As with the connectivity features, night-time activity is strongly linked to HIV and  higher activity implies higher prevalence rates. This is also confirmed by the contribution plots for the $4^{th}$- and $5^{th}$-ranked feature that encompass activity during weekday nights, between 1 AM and 2 AM and weekend nights, between 4 AM and 5 AM. On the contrary, the feature ranked $6^{th}$, which refers to early morning activity (07-08 AM) has a negative slope.

The presented contribution analysis uncovers what the trained models learned from the data. All features work in synergy to provide the prediction of the HIV prevalence. Nevertheless this method helps us to identify the subset of stronger factors. The resulting plots can be used to create new hypotheses in epidemiology, when disease distribution and spread are concerned, and, subsequently, to quantify the risk of increase in the prevalence of HIV.

\section*{Discussion}

Using mobile phone data, that can unveil patterns of human interactions and mobility, is gaining increased attention in epidemiology. In the study presented here, we placed the mobile phone data in the context of a generalized HIV epidemic. Row data was processed in the search for patterns that could explain the spatial variation in disease prevalence. We discovered that strong ties and hubs in the communication align with HIV hot spots. Strong ties created by user mobility revealed pathways that connect regions with higher prevalence, and Abidjan - the city most severely affected by HIV - emerged as the center of migrations. 

Next, we focused on extracting features related to the connectivity and mobility of users at the level of spatial units -- departments -- that could be used to predict HIV prevalence. Several regression methods were used to address that task, and  the results obtained on a subset of departments, for which good estimates of HIV prevalence exist, are promising and can lead to generation of new hypotheses. The initial set of 224 features was reduced using a recursive feature elimination procedure, allowing us to identify features with the largest impact on prediction. It turned out that night-time connectivity and activity, the spatial area covered by users and overall migrations are strongly linked to HIV prevalence. Models built on spatial features (gyration, area, perimeter of convex hull, diameter and distance) exhibit high predictive power ($\rho = 0.753,~RRMSE = 0.294$). Future work should include a detailed analysis of spatio-temporal dynamics of human motion in the context of primary and subsidiary habitats\cite{Bagrow-2012}, where the first denote frequently visited locations during typical daily activities and the second capture additional travel. 

The limitations of our study arise from spatial and temporal scale of data. On one side, HIV data is limited by the measurement strategy of DHS, UNAIDS or other relevant entities. The quality and spatial resolution of such data are determined by the sampling design - frequency and distribution of measurements. The variability in HIV prevalence across Ivory Coast is certainly higher than one modeled on the department level, but we lacked more precise measurements to account for it better. The time resolution is even scarcer. HIV measurement campaigns are organized only once every few years (for Ivory Coast 2012, 2005, 2001). Our findings linked aggregated behavioral patterns to HIV prevalence rates, but discovered correlations do not imply causation. To explore causation, we would need more estimates on changes of HIV prevalence during time. This could soon be overcome by a new device that easily connects to a smartphone\cite{Laksanasopin-2015}. The device performs the ELISA test and discovers disease markers from a tiny drop of blood, taken from a finger, in just 15 minutes. This approach has a high acceptance rate among population and will enable large--scale screening. On the other hand, the spatial resolution of mobile phone data is restricted by the distribution of carrier's antennas and the time resolution is conditioned by users' phone activity (calls or massages). But the major constraint on using mobile phone data are the privacy concerns\cite{De-2013}. Beside the mandatory user anonymization, mobile phone data are, usually, further spatially and/or temporally aggregated, or a part of information is removed. For example, the antennas are aggregated at the level of larger geographical units, time is expressed in hourly intervals, and communication graphs at the level of users are detached from any spatial information. In D4D data sources, mobile phone data sets are temporally aggregated to one-hour time slots, with preserved spatial resolution of 1250 antennas or spatially restricted to 255 sub-prefectures, but without time aggregation. Even with data aggregation, mobile phone data is still quite a richer source of information when compared to HIV estimates that are available across 50 departments. Only in the case of individual communication graphs (D4D SET4) where spatial information is completely removed, we lose any chance to link it with the HIV distribution. Those communication graphs, if geographically determined, would be an immense source of information for uncovering the connectivity at a more detailed scale. If such data becomes available in a privacy-acceptable form, further progress in the domain of modeling the spread of communicable diseases~\cite{Bian-2013} will be enabled.

In summary, our study showed how raw real--world data can be used for significant knowledge extraction. We believe that our work, a first attempt to link mobile phone data and HIV epidemiology, lays a foundation for further research into ways to explain the heterogeneity of HIV and build predictive tools aimed at advancing public--health campaigns and decision making for HIV interventions. Together with other "big data" approaches to HIV epidemiology\cite{Young-2015} that rely on Twitter data\cite{Young-2014} and social networks\cite{Young-2012, Young-2013} our work fits well into the wider initiative of digital epidemiology\cite{Salathe-2012}.

\section*{Methods}

\subsection*{Data sources}
\textit{Population data:} We used the data set available on the AfriPop website: \url{www.afripop.org}, which contains full details on population distribution, summarized on the country level. The authors developed a new high resolution population distribution data set for Africa and analyzed rural accessibility to population centers. Contemporary population data was combined with detailed satellite-derived settlement extents to map population distribution across Africa at a finer spatial resolution\cite{Linard-2012}.

\noindent\textit{HIV data:} Demographic and Health Surveys (DHS) provides data about the health status of countries. We used data collected in the survey conducted during 2011 and 2012 \cite{website-dhs}. This data provides estimates for ten administrative regions of Ivory Coast. The results of the estimation are shown in Fig. \ref{fig:hiv-distribution} (a).

\noindent\textit{D4D data:} Mobile phone data sets originate from the Orange service provider in Ivory Coast  collected in five month period (December 1, 2011 - April 28, 2012) and are further processed into four different D4D sets. Two of these were used in our study: SET1 and SET3. 
SET1 contains the antenna-to-antenna communication traffic flow of five million Orange costumers aggregated to hourly intervals. Each record contains the originating and terminating antennas of calls, the number of calls and overall duration. SET2 observes users in consecutive two-week periods, which do not significantly influence HIV transmission patterns. On the other hand, insight into the long term mobility (5 months long observation period) is possible trough SET3. Spatial resolution in this set is reduced from towers to sub-prefectures (255 spatial units). A record in this set contains the user id, time stamp and sub-prefecture ids. Although SET4 provides connectivity at the level of single users and could be very informative for HIV epidemiology, it lacks spatial information. User IDs cannot be related to the IDs in the second or third set and therefore we were not able to approximate their home locations. 

\subsection*{Estimates on HIV prevalence at the level of departments}
National estimates on HIV prevalence hide the heterogeneity that exists within the country. To unveil subnational prevalence rates, a recently proposed method - prevR  \cite{Larmarange-2011} relies on an estimation function and DHS measurements to generate a surface of HIV prevalence. Estimations are based on Gaussian kernel density functions with adaptive bandwidths. An estimate on HIV prevalence in a spatial point $(x,y)$ is determined by Eq. 1. 

\begin{equation}
\centering
\quad \quad \quad \quad \quad \quad \quad \quad \quad \quad \quad \quad \quad \quad \quad \quad prev(x,y) = \sum_{i}^{n}\frac{1}{{h_{i}}^{2}}K\left( \frac{d_{i}}{h_{i}}\right) 
\end{equation}

\noindent where, n is the number of samples, $d_i$ the geometrical distance between sample $i$ and point $(x,y)$, $K$ is the kernel function and $h_i$ the bandwidth used for sample $i$. Additionally, an indicator of the quality of the estimates was assigned to each department, based on the survey sampling size \cite{Larmarange-2014}. Some estimates are very uncertain and should be interpreted with caution. See supplement Table S1 for estimated values and quality indicators. 

\subsection*{Strong ties identification}
Ties among sub-prefectures are expressed by communication and mobility flows. To categorize  those connectivity ties as strong or weak we adopted the approach from \cite{Phithakkitnukoon-2012} where Eq. 2 is used to calculate the strength of ties.

\begin{equation}
\quad \quad \quad \quad \quad \quad \quad \quad \quad \quad \quad \quad \quad \quad \quad \quad \quad s(i) = \dfrac{c(i)}{\frac{1}{N} \sum\limits_{i=1}^N c(i)}
\end{equation}

\noindent where $i$ is index of department, $s(i)$ is the strength of tie $i$, and $c(i)$ corresponds to the number of calls or mobilities to department $i$. Ties with $s(i)<1$ are classified as weak ties, and those with $s(i) \geq 1$ as strong ones.

\subsection*{Ridge regression}
Ridge regression is a variant of ordinary multiple linear regression whose goal is to circumvent the problem of instability, arising, among other, from co-linearity of the predictor variables. It works with the original variables and tries to minimize penalized sum of squares. Like the ordinary least squares, ridge regression includes all predictor variables, but typically with smaller coefficients, depending upon the value of the complexity parameter $\lambda$. The selection of the ridge parameter $\lambda$ plays an important role; it multiplies the ridge penalty and thus controls the strength of the shrinkage of coefficients toward zero\cite{El-2011}. The value of $\lambda$ is estimated though leave-one-out validation.

\subsection*{Support vector regression}
Support vector machines are a set of supervised learning methods used for classification and regression analysis. A version of SVM for regression analysis is the Support Vector Regression (SVR)~\cite{Gunn-1998}. SVR searches for the optimal repression function, but allows a tolerance margin ($\varepsilon$), creating a tube around the regression function where errors in predictions on training data are ignored. The method also includes a regularization parameter in the form of a cost parameter ($C$), that penalizes the training errors outside the tube. In our experiments we used a linear kernel, the default $\varepsilon =0.1$, while the value of $C$ was estimated though leave-one-out validation.

\subsection*{Recursive Feature Elimination}
Recursive Feature Elimination (RFE) is a greedy method for selecting a defined number of features. It starts from the initial set of features and builds a model (in our case SVM or Ridge), assigns weights to each feature based on estimate from the predictive model, eliminates the lowest ranked feature and then recursively repeats this procedure on the remaining set of features until it reaches the desired number of features. The output is a top-ranked feature subset obtained through this recursive procedure\cite{Guyon-2002}.

\bibliography{references}

\section*{Acknowledgements}

We would like to thank the operator France Telecom-Orange and the organizers of the Data for Development Challenge for sharing the D4D data sets, as well as Joseph Larmarange from IRD (Institut de recherche pour le développement), France, for providing us with preliminary results on HIV estimates over the 50 departments of Ivory Coast. This work was partly supported by Serbian Ministry of Education and Science (Project III 43002) and by European Commission (FP7 InnoSense project, ref. no: 316191).

\section*{Author contributions}

S.B., K.G., D.\'C. and V.C. designed the research. S.B. and K.G. conducted the experiments. All authors analysed the results and participated in the writing of the manuscript  

\section*{Additional information}

Competing financial interests: The authors declare no competing financial interests.

\end{document}